\documentclass[aps,twocolumn,showpacs,byrevtex,prl,reprint]{revtex4-1}

\usepackage{graphicx}
\usepackage{dcolumn}
\usepackage{bm}
\usepackage{rotating}
\usepackage{epstopdf}
\usepackage{color}
\usepackage{verbatim} 
\usepackage{multirow}
\usepackage[abs]{overpic}
\usepackage{amsmath}
\usepackage{amssymb}
\usepackage{subfigure}
\usepackage{xspace}
\RequirePackage{lineno}

\usepackage[colorlinks,
            linkcolor=blue,
            anchorcolor=blue,
            citecolor=blue]{hyperref}

\newcommand{\PreserveBackslash}[1]{\let\temp=\\#1\let\\=\temp}
\newcolumntype{C}[1]{>{\PreserveBackslash\centering}p{#1}}
\newcolumntype{R}[1]{>{\PreserveBackslash\raggedleft}p{#1}}
\newcolumntype{L}[1]{>{\PreserveBackslash\raggedright}p{#1}}

\newcommand{\EE}{e^+e^-}

\newcommand{\too}{\rightarrow}

\begin{document}
\graphicspath{{figure/}}
\DeclareGraphicsExtensions{.eps,.png,.ps}
\title{\boldmath Observation of structures in the processes $\EE\too\omega\chi_{c1}$ and $\omega\chi_{c2}$}
\author{
  \begin{small}
    \begin{center}
      M.~Ablikim$^{1}$, M.~N.~Achasov$^{5,b}$, P.~Adlarson$^{75}$, X.~C.~Ai$^{81}$, R.~Aliberti$^{36}$, A.~Amoroso$^{74A,74C}$, M.~R.~An$^{40}$, Q.~An$^{71,58}$, Y.~Bai$^{57}$, O.~Bakina$^{37}$
% ...
\\
\vspace{0.2cm}
(BESIII Collaboration)\\
\vspace{0.2cm} {\it
$^{1}$ Institute of High Energy Physics, Beijing 100049, People's Republic of China\\
% ...
$^{5}$ Budker Institute of Nuclear Physics SB RAS (BINP), Novosibirsk 630090, Russia\\
% ...
$^{36}$ Johannes Gutenberg University of Mainz, Johann-Joachim-Becher-Weg 45, D-55099 Mainz, Germany\\
$^{37}$ Joint Institute for Nuclear Research, 141980 Dubna, Moscow region, Russia\\
% ...
$^{40}$ Liaoning Normal University, Dalian 116029, People's Republic of China\\
% ...
$^{57}$ Southeast University, Nanjing 211100, People's Republic of China\\
$^{58}$ State Key Laboratory of Particle Detection and Electronics, Beijing 100049, Hefei 230026, People's Republic of China\\
% ...
$^{71}$ University of Science and Technology of China, Hefei 230026, People's Republic of China\\
% ...
$^{74}$ University of Turin and INFN, (A)University of Turin, I-10125, Turin, Italy; (B)University of Eastern Piedmont, I-15121, Alessandria, Italy; (C)INFN, I-10125, Turin, Italy\\
$^{75}$ Uppsala University, Box 516, SE-75120 Uppsala, Sweden\\
% ...
$^{81}$ Zhengzhou University, Zhengzhou 450001, People's Republic of China\\
\vspace{0.2cm}
% ...
$^{b}$ Also at the Novosibirsk State University, Novosibirsk, 630090, Russia\\
% ...
      }\end{center}
    \vspace{0.4cm}
\end{small}
}
\affiliation{}


\begin{abstract}

We present measurements of the Born cross sections for the processes $\EE\too\omega\chi_{c1}$ and $\omega\chi_{c2}$ at center-of-mass energies $\sqrt{s}$ from 4.308 to 4.951~GeV. The measurements are performed with data samples corresponding to an integrated luminosity of 11.0$~\rm{fb}^{-1}$ collected with the BESIII detector operating at the BEPCII storage ring. Assuming the $\EE\too\omega\chi_{c2}$ signals come from a single resonance, the mass and width are determined to be $M=(4413.6\pm9.0\pm0.8)$~MeV/$c^2$ and $\Gamma=(110.5\pm15.0\pm2.9)$~MeV, respectively, which is consistent with the parameters of the well-established resonance $\psi(4415)$. In addition, we also use one single resonance to describe the $\EE\too\omega\chi_{c1}$ lineshape, and determine the mass and width to be $M=(4544.2\pm18.7\pm1.7)$~MeV/$c^2$ and $\Gamma=(116.1\pm33.5\pm1.7)$~MeV, respectively. The structure of this lineshape, observed for the first time, requires further understanding.

\end{abstract}


\maketitle

In the last decades, charmonium physics has gained great interest from both theory and experiment, stimulated by the observations of charmonium-like states, such as $X(3872)$, $Y(4230)$ and $Z_{c}(3900)$~\cite{X3872, Y4260, Zc3900}. These states, comprising of charm-anticharm ($c\bar{c}$) quark pairs, do not fit in the conventional charmonium spectroscopy, and could be exotic states~\cite{theory11, theory22, theory33, theory44}. Above the open-charm threshold, besides the three well-known vector structures observed in the inclusive hadronic cross section, i.e., $\psi(4040)$, $\psi(4160)$, and $\psi(4415)$~\cite{pdg}, a few vector $Y$ states are also observed, namely $Y(4230)$, $Y(4360)$, $Y(4500)$, $Y(4660)$ in the hidden-charm final states~\cite{Y4260, Y4360, Y4500, Y4660} and $Y(4790)$ in the $D_{s}^{*+}D_{s}^{*-}$ final state~\cite{Y4790}. The overpopulation of observed vector states compared to the expected charmonium states in this energy region implies the presence of new formation mechanisms. Some theoretical models interpret these as charmonium hybrids or tetraquark states~\cite{theory22, theory33, theory44}. More experimental measurements are needed to understand the nature of these states.

In addition, the conventional vector charmonium states above the open-charm threshold have not been understood well yet. Besides the inclusive hadronic final states, they are seldom observed in exclusive final states~\cite{pdg}. Especially the $\psi(4415)$ has never been observed in any hidden-charm decay~\cite{pdg}. Therefore, a search for new decays is essential to enhance our understanding of these states.

The BESIII Collaboration has reported the first observation of the process $\EE\too\omega\chi_{c1}$ at the center-of-mass energy $\sqrt{s}=4.600$~GeV and $\EE\too\omega\chi_{c2}$ at $\sqrt{s}=4.420$~GeV~\cite{omegachic12}. The $\sqrt{s}$-dependent lineshapes of their cross sections, however, is still absent, a key information which could help us clarify the sources of these signals and gain further insights into these observed vector states. Since then, the BESIII Collaboration has collected more data in a wider center-of-mass energy range, allowing for the determination of the $\EE\too\omega\chi_{c1,2}$ lineshape with high precision.

In this Letter, we report studies of the processes $\EE\too\omega\chi_{c1,2}$ based on the $\EE$ annihilation data taken at $\sqrt{s}=4.308$ to 4.951~GeV in 2013~---~2014 and 2019~---~2021, where $\chi_{c1,2}$ are reconstructed via the decay modes $\chi_{c1,2}\too\gamma J/\psi$, $J/\psi\too\ell^{+}\ell^{-}$ $(\ell=e,\mu)$, and $\omega$ is reconstructed via its decay $\omega\too\pi^{+}\pi^{-}\pi^{0}$, $\pi^0\too\gamma\gamma$.

The BESIII detector is a magnetic spectrometer~\cite{besiii} located at the Beijing Electron Positron Collider (BEPCII)~\cite{bepcii}. The cylindrical core of the BESIII detector consists of a helium-based multilayer drift chamber (MDC), a plastic scintillator time-of-flight system (TOF), and a CsI (Tl) electromagnetic calorimeter (EMC), which are all enclosed in a superconducting solenoidal magnet, providing a 1.0~T magnetic field. The solenoid is supported by an octagonal flux-return yoke with resistive plate chamber muon identifier modules interleaved with steel. The acceptance of charged particles and photons is 93\% over the $4\pi$ solid angle. The charged-particle momentum resolution at $1~{\rm GeV}/c$ is $0.5\%$, and the $\textrm{d}E/\textrm{d}x$ resolution is $6\%$ for the electrons from Bhabha scattering. The EMC measures photon energies with a resolution of $2.5\%$ ($5\%$) at $1$~GeV in the barrel (end cap) region. The time resolution of the TOF barrel section is 68~ps, while that of the end cap section is 110~ps. The end cap TOF system was upgraded in 2015 with multi-gap resistive plate chamber technology, providing a time resolution of 60~ps~\cite{etof}; this improvement benefits about 70\% of the data sample used in this work.

This analysis is performed based on the data samples collected at twenty-five energy points in the range of $\sqrt{s}=4.308$ to 4.951~GeV. The center-of-mass energy and the integrated luminosity of each sample are measured with the di-muon and Bhabha processes, respectively~\cite{luminosity1, luminosity2, luminosity3}, as listed in the Supplemental Material~\cite{supplement}. A {\sc geant4} based Monte Carlo (MC) simulation is implemented to optimize the signal selection criteria, extract the efficiencies, and study the backgrounds. The $\EE\too\omega\chi_{c1,2}$ signal samples at each energy point are simulated according to the phase space model with the Initial State Radiation (ISR) effect taken into account.

The candidate events must comprise four good charged tracks with zero net charge and at least three photons because the final states of the signal processes include $3\gamma\pi^{+}\pi^{-}\ell^+\ell^-$ $(\ell=e,\mu)$. Here, each good charged track needs to satisfy the following criteria: the distance of closest approach to the interaction point is within $10$~cm along the beam direction and 1~cm in the plane perpendicular to the beam direction. In addition, the polar angles ($\theta$) of the tracks must be within the fiducial volume of the MDC $(|\cos\theta|<0.93)$. Photon candidates are reconstructed from isolated showers in the EMC at least $10^\circ$ away from the nearest charged tracks. The photon energy is required to be at least 25~MeV in the barrel region $(|\cos\theta|<0.80)$ or 50~MeV in the end cap region $(0.86<|\cos\theta|<0.92)$. To suppress electronic noise and energy depositions unrelated to the event, the EMC cluster timing from the reconstructed event start time is further required to satisfy $0\leq t \leq 700$~ns.

The tracks with momenta higher than 1~GeV/$c$ are identified as the leptons from $J/\psi$, otherwise they are treated as the pions from the $\omega$ decay. Furthermore, electron candidates need to deposit an energy larger than 1~GeV in the EMC, muon candidates less than 0.4~GeV. In order to further suppress the backgrounds and improve the mass resolutions, a five constraint (5C) kinematic fit is implemented by constraining the total four-momentum of the reconstructed particles to the total four-momentum of the colliding beams, and the invariant mass of two photons to the $\pi^{0}$ nominal mass. If more than one combination is found among the final state particles, the one with the least $\chi^{2}_{\rm{5C}}$ of the kinematic fit is chosen. The $\chi^{2}_{\rm{5C}}$ is required to be less than 60.

\begin{figure}[htbp]
\begin{center}
\begin{overpic}[width=0.22\textwidth]{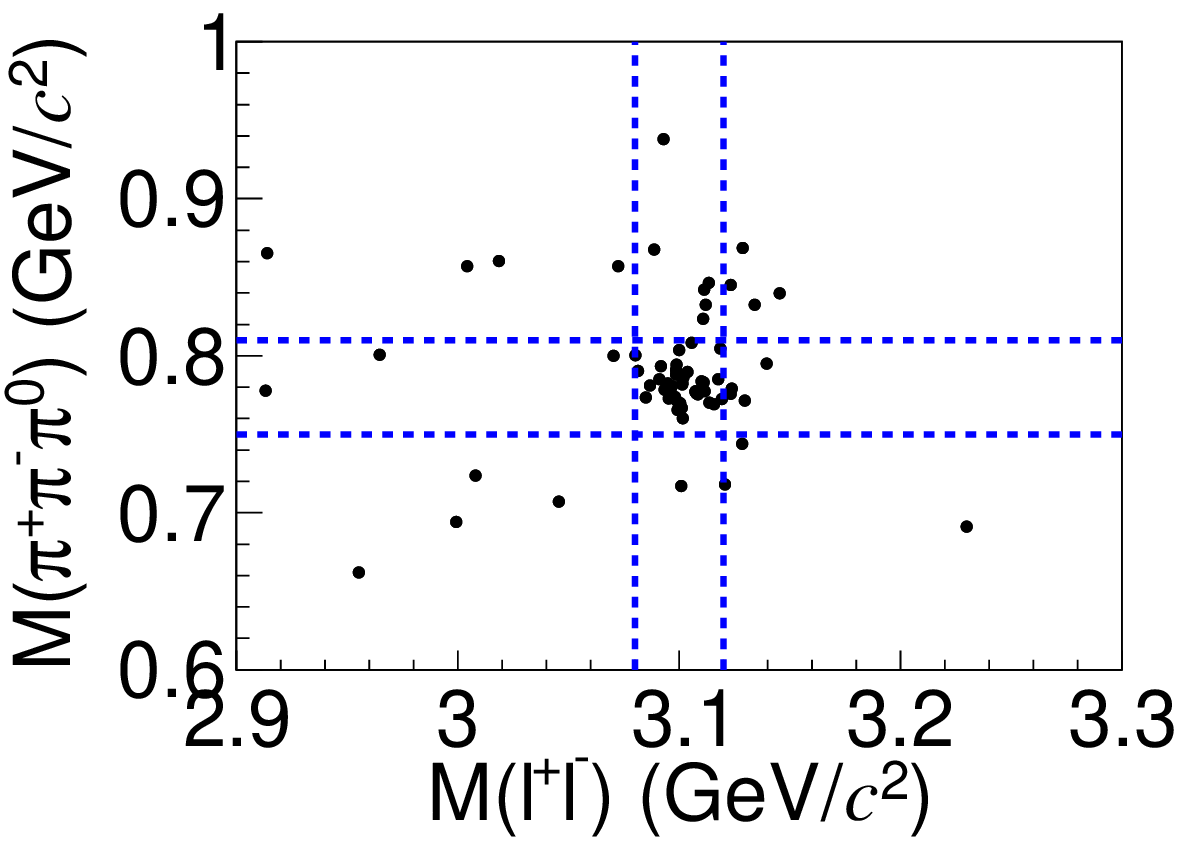}
\put(91,65){(a)}
\end{overpic}
\begin{overpic}[width=0.22\textwidth]{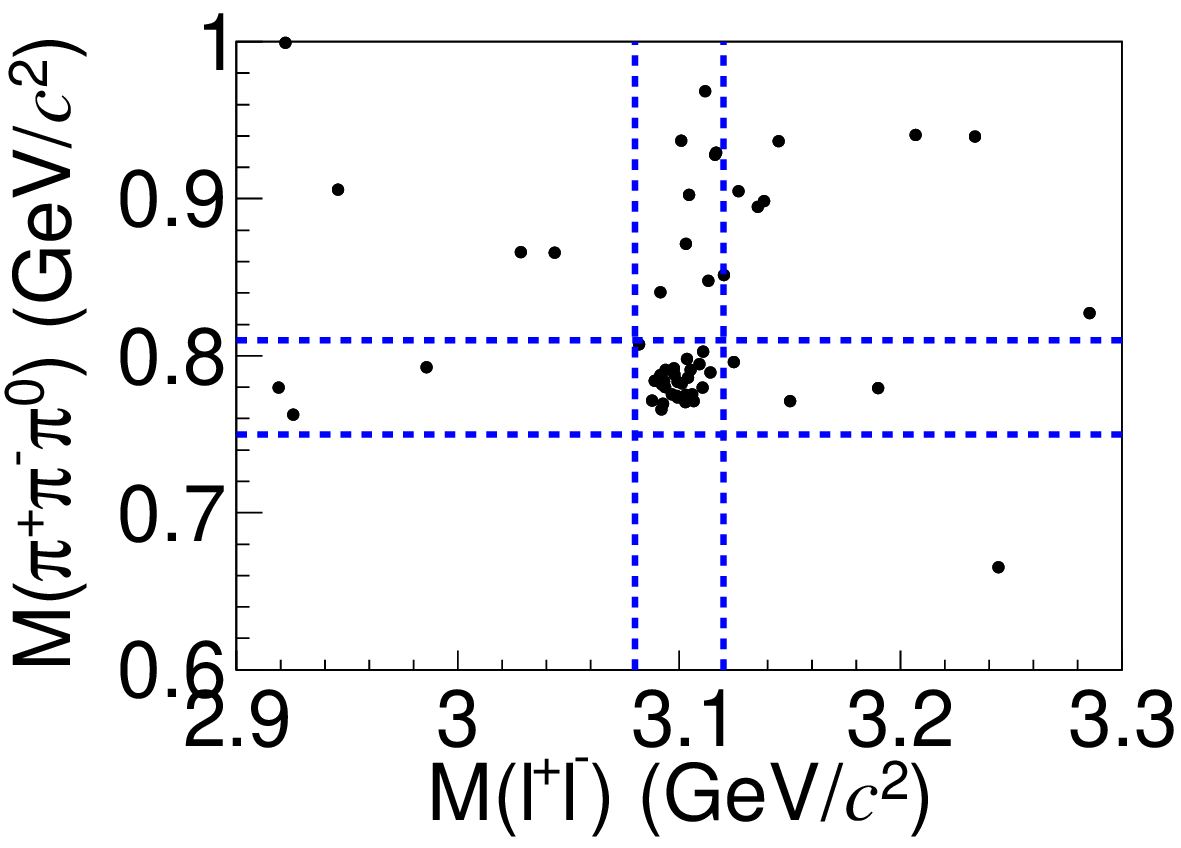}
\put(91,65){(b)}
\end{overpic}
\begin{overpic}[width=0.22\textwidth]{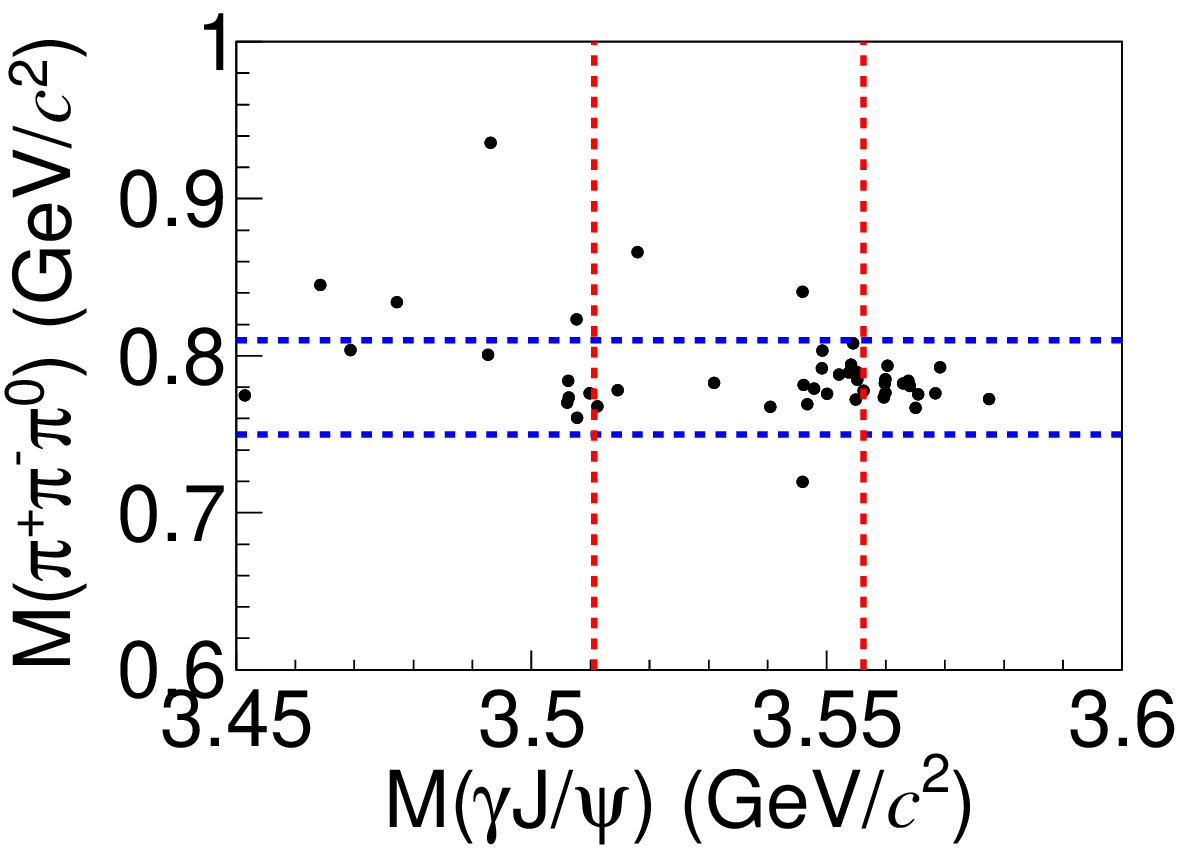}
\put(91,65){(c)}
\end{overpic}
\begin{overpic}[width=0.22\textwidth]{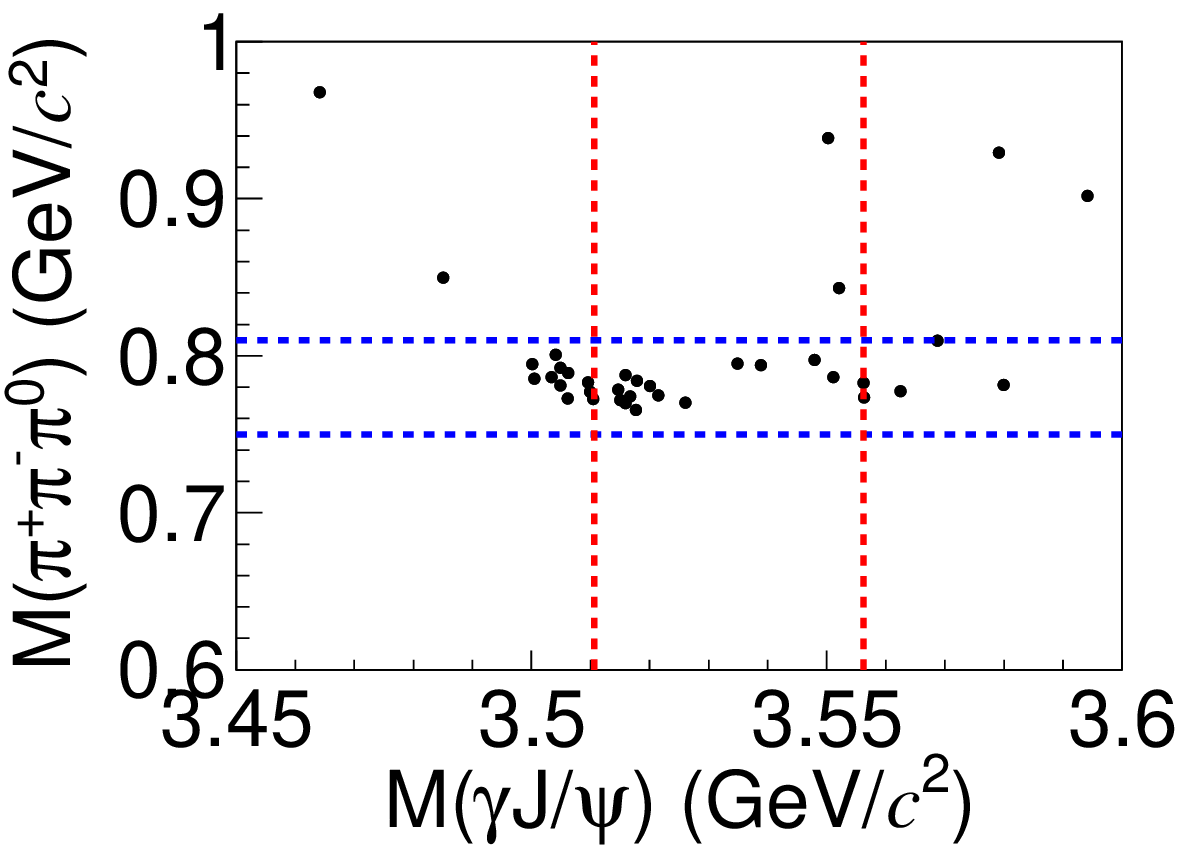}
\put(91,65){(d)}
\end{overpic}
\caption{Distributions of $M(\pi^{+}\pi^{-}\pi^{0})$ versus $M(\ell^{+}\ell^{-})$ for data at $\sqrt{s}=4.436$~GeV (a) and 4.600~GeV (b), and distributions of $M(\pi^{+}\pi^{-}\pi^{0})$ versus $M(\gamma J/\psi)$ for data at $\sqrt{s}=4.436$~GeV (c) and 4.600~GeV (d). The blue dashed lines mark the signal regions of $\omega$ and $J/\psi$, and the red dashed lines mark the nominal masses of $\chi_{c1,2}$.}
\label{fig:scatter}
\end{center}
\end{figure}

The results at $\sqrt{s}=4.436$~GeV and 4.600~GeV are shown and discussed as examples in the main text due to their high signal significances. With all the event selection criteria imposed, the distributions of $M(\pi^{+}\pi^{-}\pi^{0})$ versus $M(\ell^{+}\ell^{-})$ of the candidate events are shown in Fig.~\ref{fig:scatter}. Clear event accumulations appear in the $\omega$ and $J/\psi$ signal regions, which are defined as [0.75, 0.81]~GeV/$c^2$ and [3.08, 3.12]~GeV/$c^2$, respectively. The $\omega$ and $J/\psi$ mass windows are about three times the width of signal MC shape. The efficiency of the mass windows varies from $82\%$ to $91\%$ depending on the collision energy. The distributions of $M(\pi^{+}\pi^{-}\pi^{0})$ versus $M(\gamma J/\psi)$ after the $J/\psi$ mass window requirement are also shown in Fig.~\ref{fig:scatter}. Significant $\omega\chi_{c2}$ signals are observed at $\sqrt{s}=4.436$~GeV, the $\omega\chi_{c1}$ signals at $\sqrt{s}=4.600$~GeV.

\begin{figure}[htbp]
\begin{center}
\includegraphics[width=0.23\textwidth]{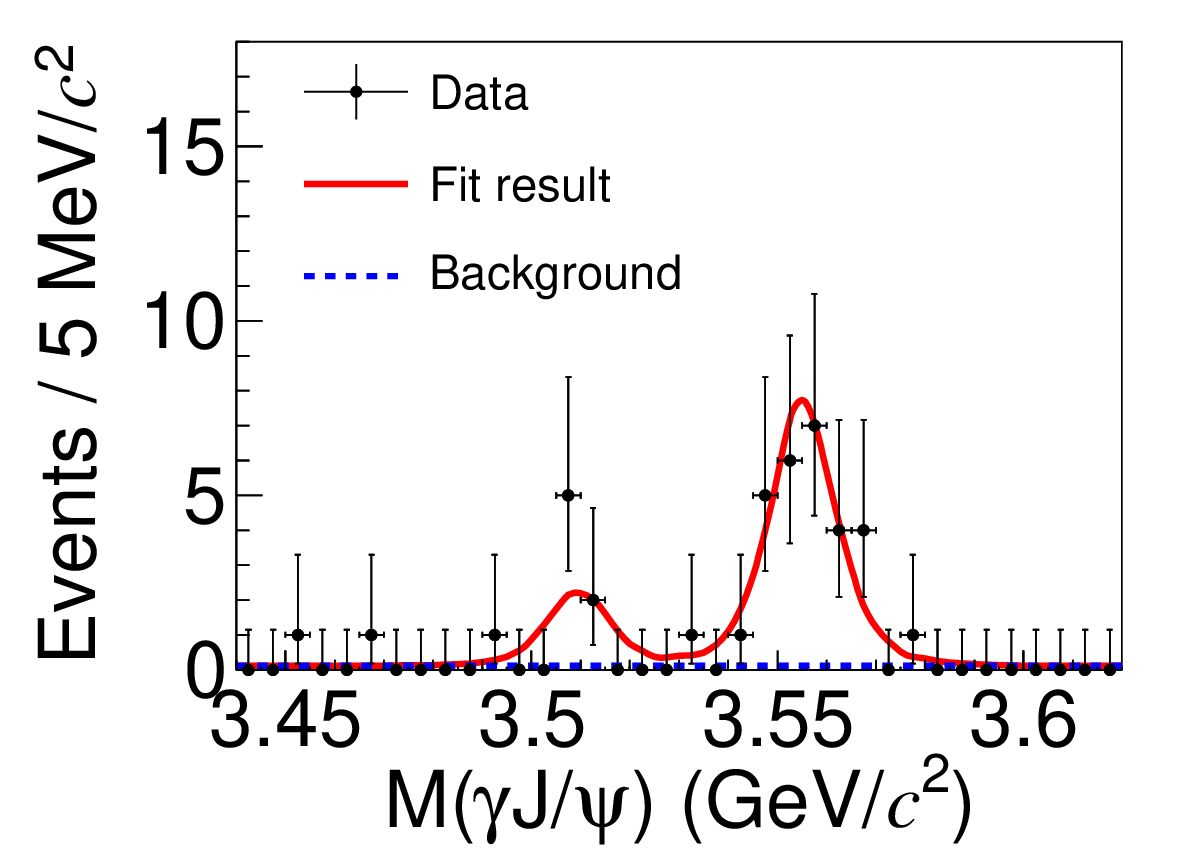}
\includegraphics[width=0.23\textwidth]{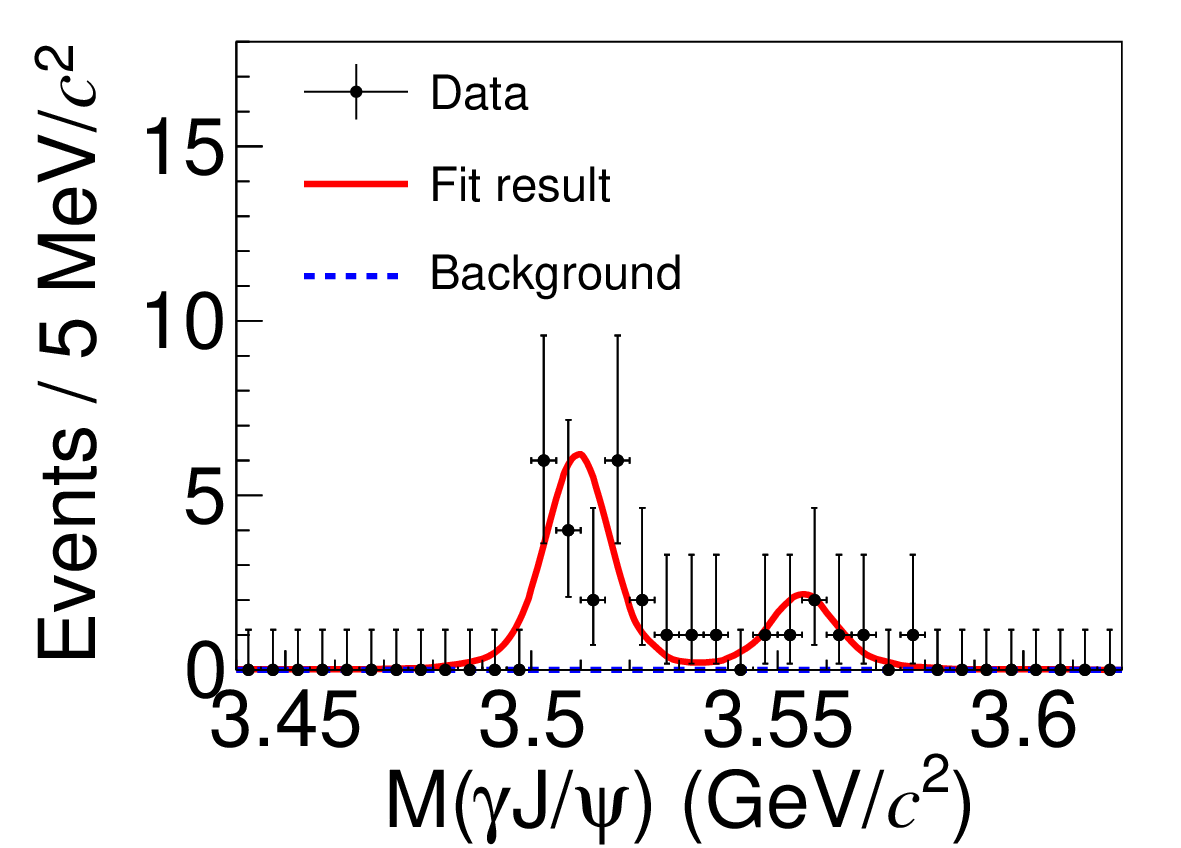}
\caption{Fits to the $M(\gamma J/\psi)$ distributions at $\sqrt{s}=4.436$~GeV (left) and 4.600~GeV (right), respectively.}
\label{fig:10}
\end{center}
\end{figure}

After performing the selection of the $\omega$ signal, Fig.~\ref{fig:10} shows the corresponding one-dimensional projections on the $M(\gamma J/\psi)$ distribution which are used to extract the signal yields. The backgrounds are studied using the $\omega$-$J/\psi$ two-dimensional sideband regions defined as [0.66, 0.72]~GeV/$c^2$ and [0.84, 0.90]~GeV/$c^2$ for the $\omega$ and [3.00, 3.06]~GeV/$c^2$ and [3.14, 3.20]~GeV/$c^2$ for the $J/\psi$. No significant peaking background is observed.

For the energy points where the signal is observed with significance above $3\sigma$, an unbinned maximum likelihood fit is performed on the $M(\gamma J/\psi)$ spectrum of selected events to determine the $\chi_{c1}$ and $\chi_{c2}$ yields. The fit function is a sum of the MC-determined $\chi_{c1}$ and $\chi_{c2}$ shapes and a flat background. Fig.~\ref{fig:10} shows the fit results at $\sqrt{s} = 4.436$~GeV and 4.600~GeV. The goodness of the fit $\chi^2/ndf$ is $29.3/33$ at $\sqrt{s} = 4.436$ and $27.1/33$ at 4.600~GeV, where $ndf$ denotes the number of degrees of freedom. For the energy points with a signal significance less than $3\sigma$, the signal yields are obtained by directly counting the events in the $\chi_{c1}$ and $\chi_{c2}$ signal regions which are defined as [3.49, 3.53]~GeV/$c^2$ and [3.536, 3.576]~GeV/$c^2$, respectively. The background is estimated in the $\chi_{c1,2}$ sideband in [3.43, 3.47]~GeV/$c^2$ and then subtracted. The results at all the energy points used in this work are summarized in the Supplemental Material~\cite{supplement}.

The Born cross section at each energy point is calculated by
\begin{linenomath*}
\begin{equation}
    \sigma^{B}(e^+e^-\too\omega\chi_{c1,2}) = \frac{N^{\rm{sig}}}{\mathcal{L}_{\rm{int}}\epsilon(1+\delta)\frac{1}{|1-\Pi|^{2}}(\mathcal{B}_{e} + \mathcal{B}_{\mu})\mathcal{B}_{1}},
\end{equation}
\end{linenomath*}
where $N^{\rm{sig}}$ is the signal yield, $\mathcal{L}_{\rm{int}}$ is the integrated luminosity, $\epsilon$ is the selection efficiency obtained with the signal MC, $\mathcal{B}_{e}$ is the branching fraction $\mathcal{B}(J/\psi\too\EE)$, $\mathcal{B}_{\mu}$ is $\mathcal{B}(J/\psi\too\mu^{+}\mu^{-})$, $\mathcal{B}_{1}$ is $\mathcal{B}(\chi_{c1,2}\too\gamma J/\psi)\times\mathcal{B}(\omega\too\pi^{+}\pi^{-}\pi^{0})\times\mathcal{B}(\pi^{0}\too\gamma\gamma)$, $\frac{1}{|1-\Pi|^{2}}$ is the vacuum polarization factor~\cite{vacuum}, and $(1+\delta)$ is the radiative correction factor~\cite{kkmc, QED}. We can see the cross section is obtained using selection efficiency and radiative correction factor, which, in turn, depends on the input cross section distribution in MC simulation~\cite{isrmeth}. So the iterative procedure is used to determine the Born cross section~\cite{isrmeth}. The values of the $\EE\too\omega\chi_{c1,2}$ cross sections at different energy points can be found in the Supplemental Material~\cite{supplement}.

The systematic uncertainties of the Born cross section measurements mainly come from luminosity measurement, tracking efficiency, photon detection, kinematic fit requirement, $J/\psi$ mass window, angular distribution, lineshape, fit range, signal shape, background shape, and branching fraction.

The uncertainty from the luminosity measurement is about $1.0\%$~\cite{luminosity1, luminosity2, luminosity3}. The uncertainty in tracking efficiency is obtained as $1.0\%$ per track~\cite{track1}, so a 4.0$\%$ uncertainty contributes to the final results. The uncertainty in photon reconstruction is about $1.0\%$ per photon, which is estimated with a control sample of $J/\psi\too\rho^{0}\pi^{0}$ decays~\cite{track2}.

The uncertainty caused by the 5C kinematic fit is estimated with the method of correcting the track helix parameters, where the correction factors for $\pi$, $e$ and $\mu$ are obtained by using control samples of $\EE\too\pi^{+}\pi^{-}J/\psi$, $J/\psi\too\EE$ and $\mu^{+}\mu^{-}$~\cite{helix}. The difference in detection efficiency with and without the correction is taken as the systematic uncertainty. The corresponding uncertainty from the $J/\psi$ mass window cut is estimated using the control sample $\EE\too\gamma_{\rm{ISR}}\psi(3686)$, $\psi(3686)\too\pi^{+}\pi^{-}J/\psi$, and 1.6$\%$ is assigned as the associated systematic uncertainty~\cite{jpsimasswindow}.

In order to estimate the uncertainty caused by the $\omega$ angular distribution in the signal simulation, the $\omega$ helicity angle function is set to $1\pm{\cos^{2}\theta}$ in the generator instead of the phase space model, where $\theta$ is the polar angle of $\omega$ in the $\EE$ rest frame. The efficiency difference between them divided by $\sqrt{12}$ is used as the systematic uncertainty~\cite{angular}. For the processes $\EE\too\omega\chi_{c1,2}$, single Breit-Wigner (BW) functions describe the lineshapes well. To estimate the uncertainty from the lineshapes, we change the mass and width of the BW functions by $\pm{1\sigma}$, and the maximum change of the results is regarded as the systematic uncertainty.

To examine the systematic uncertainty due to the fits to extract the $\chi_{c1,2}$ yields, we change the fit range by $\pm10$~MeV$/c^{2}$, and the maximum difference between the new and nominal results is regarded as the systematic uncertainty. In order to estimate the uncertainty from the signal shape, we use the MC-determined lineshape convolved with a Gaussian function as an alternative description, and the change to the fitted yield is regarded as the systematic uncertainty. In order to estimate the systematic uncertainty caused by the smooth background shape, which is constant in the nominal fit, we describe the background by a linear function, and the difference to the fitted signal yield is regarded as the systematic uncertainty. The uncertainty from the branching fractions quoted from the PDG~\cite{pdg} is considered as a systematic uncertainty.

All the systematic uncertainties on the $\EE\too\omega\chi_{c1,2}$ cross sections are summarized in the Supplemental Material~\cite{supplement}. The overall systematic uncertainties are obtained by adding each systematic uncertainty in quadrature under the assumption that they are independent. This results in a total systematic uncertainty of about 6\%$\sim$10\% for each energy point.

\begin{figure}[htbp]
\begin{center}
\includegraphics[width=0.23\textwidth]{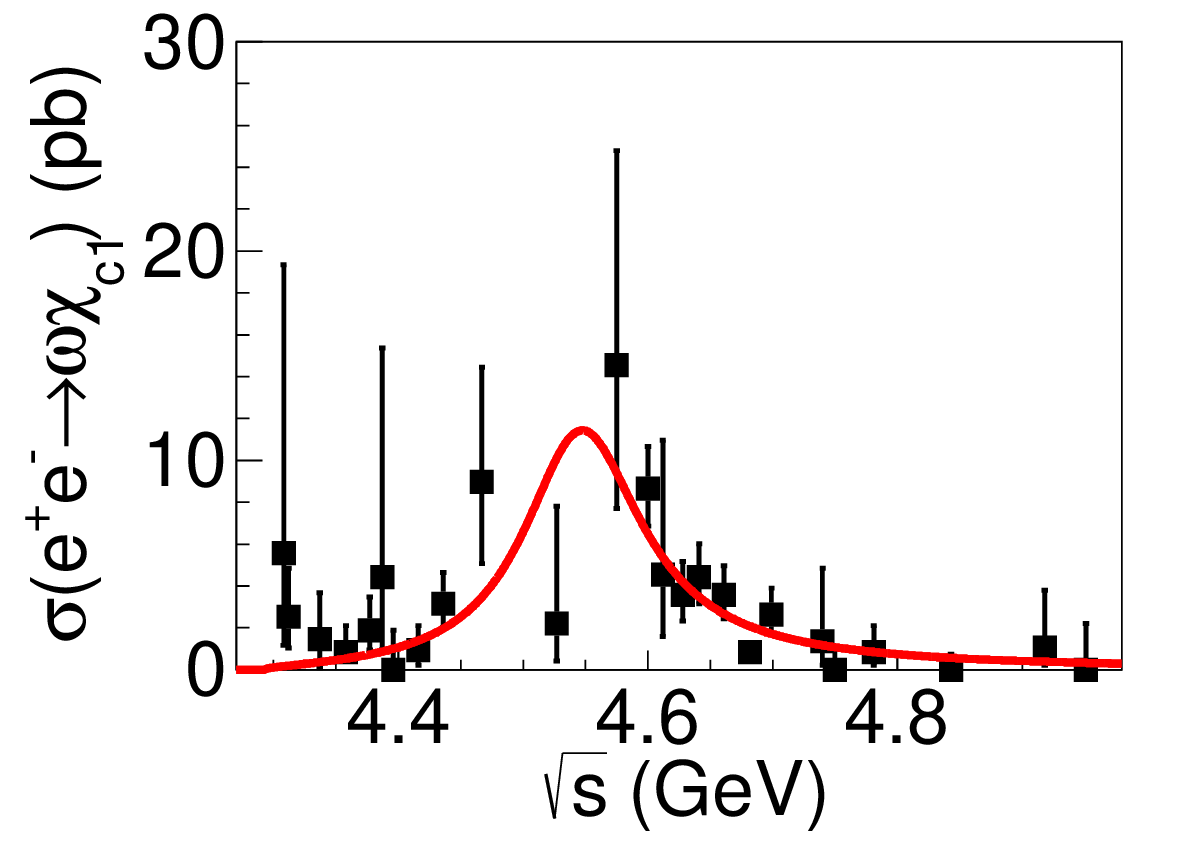}
\includegraphics[width=0.23\textwidth]{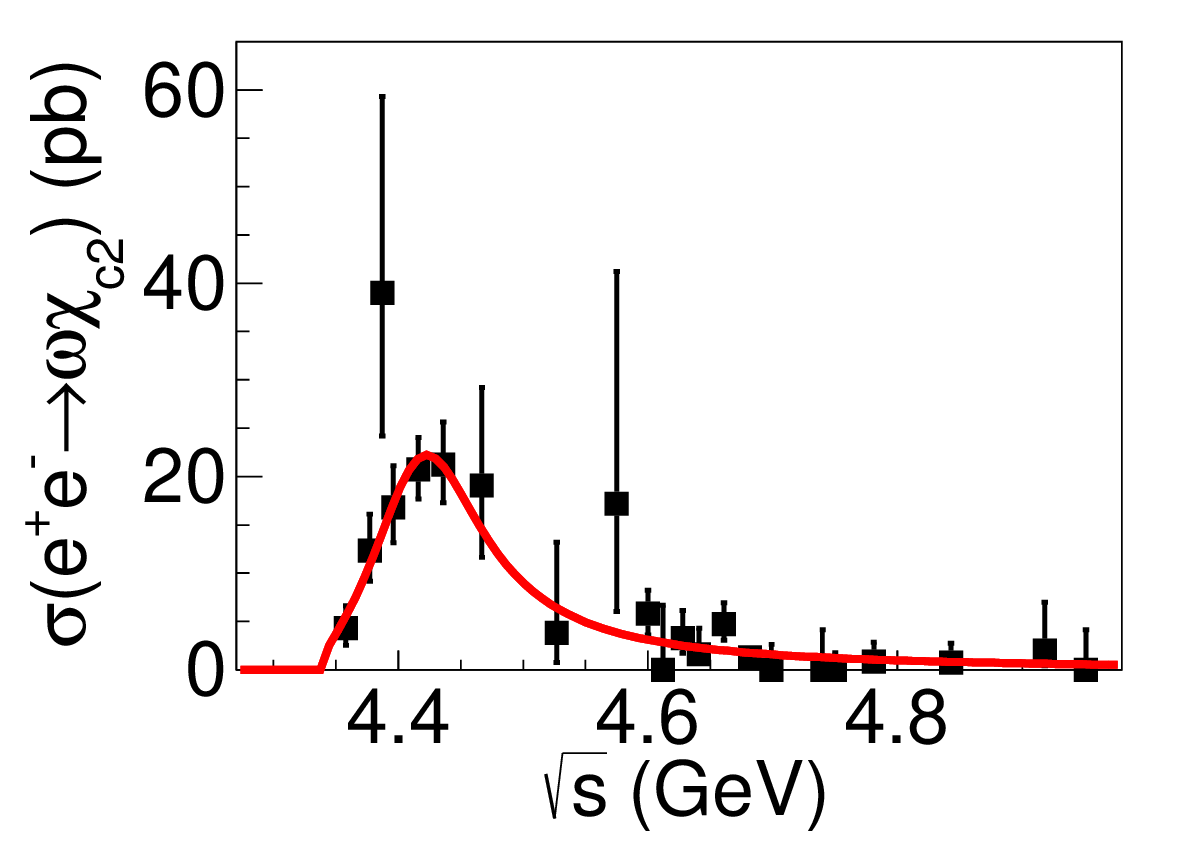}
\caption{Fits to the cross sections of $\EE\too\omega\chi_{c1}$ (left) and $\EE\too\omega\chi_{c2}$ (right) with one single resonance.}
\label{fig:crosssection1}
\end{center}
\end{figure}

Fig.~\ref{fig:crosssection1} shows the $\EE\too\omega\chi_{c1,2}$ dressed cross sections (the cross section without the correction for vacuum polarization, $\sigma=\frac{\sigma^{B}}{|1-\Pi|^2}$) at each energy point. Assuming that the $\omega\chi_{c1}$ and $\omega\chi_{c2}$ signals each come from one single resonance, we fit these cross sections with a maximum likelihood method. In the fit, the structure is described by a BW function,

\begin{linenomath*}
\begin{equation}
    \sigma(\sqrt{s}) = \frac{12\pi\Gamma_{ee}\mathcal{B}(\omega\chi_{cJ})\Gamma}{(s-M^{2})^2+M^2\Gamma^2}\times\frac{\Phi(\sqrt{s})}{\Phi(M)},
\end{equation}
\end{linenomath*}
where $M$ is the mass, $\Gamma$ is the total width, $\Gamma_{ee}$ is the electronic partial width, and $\Phi(\sqrt{s})$ is the two-body phase space factor. The fit results are $M_1=(4544.2\pm18.7\pm1.7)$~MeV/$c^{2}$, $\Gamma_1=(116.1\pm33.5\pm1.7)$~MeV, $\Gamma_{ee}\mathcal{B}(\omega\chi_{c1})=(1.86\pm0.32\pm0.13)$~eV for the $\EE\too\omega\chi_{c1}$ process, and $M_2=(4413.6\pm9.0\pm0.8)$~MeV/$c^{2}$, $\Gamma_2=(110.5\pm15.0\pm2.9)$~MeV, $\Gamma_{ee}\mathcal{B}(\omega\chi_{c2})=(3.17\pm0.39\pm0.24)$~eV for the $\EE \too \omega \chi_{c2}$ process, where the first uncertainties are statistical, and the second systematic as discussed later. The fit qualities are $\chi^2/ndf=25.0/22$ for $\omega\chi_{c1}$ and $\chi^2/ndf=11.9/19$ for $\omega\chi_{c2}$, both are acceptable, implying that the $\EE \too \omega\chi_{c1,2}$ cross sections can be well described by a single BW function, respectively. With the current amount of data, it is unclear whether there are more resonances in these two lineshapes, which needs more data to confirm. We also try to fit the $\EE \too \omega\chi_{c1,2}$ cross sections using a coherent sum of BW function and phase space term, and find that the phase space term does not contribute significantly. The statistical significances of the two resonances over the phase space term are $5.9\sigma$ for $\EE\too\omega\chi_{c1}$ and $10.7\sigma$ for $\EE\too\omega\chi_{c2}$. Taking into account systematic uncertainties decreases the significance of the structure in $\EE\too\omega\chi_{c1}$ by less than $0.1\sigma$. This indicates that the structure in $\EE\too\omega\chi_{c1}$ is observed for the first time. The parameters of the structure observed in $\EE\too\omega\chi_{c2}$ are consistent with the known $\psi$(4415) parameters~\cite{pdg}.

The sources of systematic uncertainties on the resonant parameters are dominated by those due to the beam energy, BW parametrization and cross section measurement.

We conservatively take 0.8~MeV as the systematic uncertainty in the beam energy~\cite{ecms} for the mass measurements of the resonances. To estimate the uncertainty from the BW parametrization, $\Gamma$ is set to be the energy-dependent width $\Gamma(\sqrt{s})=\Gamma^{0}\frac{\sqrt{s}}{M}$, where $\Gamma^{0}$ is the nominal width of the resonance. The difference between the updated and nominal results are taken as the corresponding systematic uncertainty.

The systematic uncertainty of the cross section measurement consists of two parts: one is the uncorrelated uncertainty due to the 5C kinematic fit, lineshape and angular distribution, while the other is the common uncertainty that includes all other systematic sources mentioned above. The former part is considered by including the uncorrelated systematic uncertainty in the fit to the cross section, the change of the fitted parameter is taken as the systematic uncertainty. The latter part is common for all data points, and we simultaneously vary the cross sections at each energy point by $\pm1\sigma$ of the systematic uncertainty. The difference between the new and nominal results are taken as the systematic uncertainty. The total systematic uncertainty is obtained by adding these two items in quadrature under the assumption that they are independent.

Table~\ref{tab:sumfiterror} summarizes all the systematic uncertainties of the resonant parameters. The overall systematic uncertainties are obtained by adding all the sources of systematic uncertainties in quadrature.
\begin{table}[htbp]
\begin{center}
\caption{Systematic uncertainties on the resonant parameters. The first values in brackets are for the structure in $\EE\too\omega\chi_{c1}$, and the second for the structure in $\EE\too\omega\chi_{c2}$.}
\label{tab:sumfiterror}
\begin{tabular}{c  c c c}
  \hline
  \hline
  & $\Gamma_{ee}\mathcal{B}(\omega\chi_{cJ})$ (eV) & $M$ (MeV/$c^{2}$) & $\Gamma$ (MeV) \\
  \hline
  Beam energy     & ($-$, $-$)   & (0.8, 0.8) & ($-$, $-$) \\
  Parametrization & (0.01, 0.07) & (1.1, 0.2) & (0.2, 2.9) \\
  Cross section   & (0.13, 0.23) & (0.9, 0.1) & (1.7, 0.3) \\
  Total           & (0.13, 0.24) & (1.7, 0.8) & (1.7, 2.9) \\
  \hline
  \hline
\end{tabular}
\end{center}
\end{table}

In summary, the $\EE\too\omega\chi_{c1,2}$ Born cross sections have been measured at the center-of-mass energies from 4.308 to 4.951~GeV at the BESIII experiment. For each process, a non-trivial feature is observed in the cross section lineshape. Under the assumption that one single resonance describes the corresponding shape, the mass and width for $\EE\too\omega\chi_{c1}$ are determined to be $M=(4544.2\pm18.7\pm1.7)$~MeV/$c^2$ and $\Gamma=(116.1\pm33.5\pm1.7)$~MeV. This structure is observed for the first time with a significance of $5.8\sigma$. The mass is significantly higher compared to the structures recently observed around $4.480$~GeV/$c^2$ in $\EE\too K^+K^-J/\psi$ and $\EE\too D^{*0}D^{*-}\pi^+$~\cite{Y4500, dstardstarpi}. It is yet unclear whether or not these two states are the same based on the currently available information, and further measurements with higher precision in this energy region will be necessary. For the $\EE\too\omega\chi_{c2}$ process, the extracted parameters are $M=(4413.6\pm9.0\pm0.8)$~MeV/$c^2$ and $\Gamma=(110.5\pm15.0\pm2.9)$~MeV, suggesting that this is likely the well known $\psi(4415)$ and implying the existence of the hidden charm decay $\psi(4415)\too\omega\chi_{c2}$.

The BESIII Collaboration thanks the staff of BEPCII and the IHEP computing center for their strong support. This work is supported in part by National Key R\&D Program of China under Contracts Nos. 2020YFA0406300, 2020YFA0406400; National Natural Science Foundation of China (NSFC) under Contracts Nos. 12375071, 11635010, 11735014, 11835012, 11935015, 11935016, 11935018, 11961141012, 12022510, 12025502, 12035009, 12035013, 12061131003, 12192260, 12192261, 12192262, 12192263, 12192264, 12192265, 12221005, 12225509, 12235017; the Chinese Academy of Sciences (CAS) Large-Scale Scientific Facility Program; the CAS Center for Excellence in Particle Physics (CCEPP); Joint Large-Scale Scientific Facility Funds of the NSFC and CAS under Contract No. U1832207; CAS Key Research Program of Frontier Sciences under Contracts Nos. QYZDJ-SSW-SLH003, QYZDJ-SSW-SLH040; 100 Talents Program of CAS; The Institute of Nuclear and Particle Physics (INPAC) and Shanghai Key Laboratory for Particle Physics and Cosmology; European Union's Horizon 2020 research and innovation programme under Marie Sklodowska-Curie grant agreement under Contract No. 894790; German Research Foundation DFG under Contracts Nos. 455635585, Collaborative Research Center CRC 1044, FOR5327, GRK 2149; Istituto Nazionale di Fisica Nucleare, Italy; Ministry of Development of Turkey under Contract No. DPT2006K-120470; National Research Foundation of Korea under Contract No. NRF-2022R1A2C1092335; National Science and Technology fund of Mongolia; National Science Research and Innovation Fund (NSRF) via the Program Management Unit for Human Resources \& Institutional Development, Research and Innovation of Thailand under Contract No. B16F640076; Polish National Science Centre under Contract No. 2019/35/O/ST2/02907; The Swedish Research Council; U. S. Department of Energy under Contract No. DE-FG02-05ER41374.

\clearpage

\begin{widetext}

\textbf{\large
\boldmath Supplemental Material for ``Observation of structures in the processes $\EE\too\omega\chi_{c1}$ and $\omega\chi_{c2}$"
\unboldmath
}

\vspace{0.7cm}

The results for $\EE\too\omega\chi_{c1,2}$ cross sections at different energy points are summarized in Tables~\ref{tab:crosssectionsum1} and~\ref{tab:crosssectionsum2}, respectively. Table~\ref{tab:sumerror} summarizes all the systematic uncertainties on the $\EE\too\omega\chi_{c1,2}$ cross section measurements.

\begin{table}[htbp]
\centering
\caption{The results on Born cross sections of $\EE\too\omega\chi_{c1}$ at different energy points. The uncertainties are statistical.}
\label{tab:crosssectionsum1}
\begin{tabular}{ccccccccc}
  \hline
  \hline
  \ \ \ $\sqrt{s}$ (GeV) \ \ \ & \ \ \ $N^{\rm{sig}}$ \ \ \ & \ \ \ $\mathcal{L}_{\rm{int}}$ (pb$^{-1}$) \ \ \ & \ \ \ $\epsilon$ $(\%)$ \ \ \ & \ \ \ 1+$\delta$ \ \ \ & \ \ \ $\frac{1}{|1-\Pi|^{2}}$ \ \ \ & \ \ \ $\sigma^{B}$ (pb) \ \ \ \\
  \hline
  4.308 & $1.0_{-0.8}^{+2.5}$  & 45   & 16.35 & 0.682 & 1.052 & $5.3_{-4.2}^{+13.1}$ \\
  4.312 & $5.0_{-2.9}^{+4.6}$  & 501  & 15.54 & 0.694 & 1.052 & $2.4_{-1.4}^{+2.2}$ \\
  4.337 & $3.0_{-2.9}^{+4.6}$  & 505  & 15.51 & 0.732 & 1.051 & $1.4_{-1.3}^{+2.1}$ \\
  4.358 & $2.0_{-1.3}^{+2.8}$  & 544  & 15.39 & 0.747 & 1.051 & $0.8_{-0.5}^{+1.2}$ \\
  4.377 & $4.0_{-1.9}^{+3.4}$  & 523  & 14.73 & 0.754 & 1.052 & $1.8_{-0.9}^{+1.5}$ \\
  4.387 & $1.0_{-0.8}^{+2.5}$  & 56   & 14.92 & 0.757 & 1.052 & $4.2_{-3.3}^{+10.4}$ \\
  4.396 & $0.0_{-1.8}^{+3.7}$  & 508  & 14.44 & 0.759 & 1.052 & $0.0_{-0.9}^{+1.8}$ \\
  4.416 & $4.0_{-3.1}^{+4.8}$  & 1044 & 14.62 & 0.761 & 1.053 & $0.9_{-0.7}^{+1.1}$ \\
  4.436 & $7.1_{-2.6}^{+3.2}$  & 570  & 14.36 & 0.760 & 1.055 & $3.0_{-1.1}^{+1.4}$ \\
  4.467 & $3.9_{-1.7}^{+2.4}$  & 111  & 14.37 & 0.757 & 1.055 & $8.5_{-3.7}^{+5.2}$ \\
  4.527 & $1.0_{-0.8}^{+2.5}$  & 112  & 14.43 & 0.765 & 1.055 & $2.1_{-1.7}^{+5.3}$ \\
  4.575 & $3.0_{-1.4}^{+2.1}$  & 49   & 13.74 & 0.848 & 1.055 & $13.8_{-6.5}^{+9.7}$ \\
  4.600 & $21.8_{-4.5}^{+5.1}$ & 587  & 12.94 & 0.920 & 1.055 & $8.2_{-1.7}^{+1.9}$ \\
  4.612 & $2.0_{-1.3}^{+2.8}$  & 104  & 12.24 & 0.957 & 1.055 & $4.3_{-2.8}^{+6.1}$ \\
  4.628 & $8.0_{-2.8}^{+3.6}$  & 522  & 11.68 & 1.005 & 1.055 & $3.4_{-1.2}^{+1.5}$ \\
  4.641 & $10.4_{-3.0}^{+3.7}$ & 552  & 11.34 & 1.045 & 1.055 & $4.2_{-1.2}^{+1.5}$ \\
  4.661 & $8.1_{-2.5}^{+3.2}$  & 530  & 10.68 & 1.104 & 1.055 & $3.4_{-1.1}^{+1.3}$ \\
  4.682 & $6.0_{-2.4}^{+3.8}$  & 1669 & 10.00 & 1.164 & 1.055 & $0.8_{-0.3}^{+0.5}$ \\
  4.699 & $5.8_{-2.1}^{+2.8}$  & 536  & 9.56  & 1.211 & 1.055 & $2.5_{-0.9}^{+1.2}$ \\
  4.740 & $1.0_{-0.8}^{+2.5}$  & 164  & 9.14  & 1.318 & 1.055 & $1.3_{-1.1}^{+3.3}$ \\
  4.750 & $0.0_{-0.0}^{+1.6}$  & 367  & 8.80  & 1.342 & 1.055 & $0.0_{-0.0}^{+1.0}$ \\
  4.781 & $2.0_{-1.3}^{+2.8}$  & 513  & 8.56  & 1.416 & 1.055 & $0.8_{-0.6}^{+1.2}$ \\
  4.843 & $0.0_{-0.0}^{+1.6}$  & 527  & 7.86  & 1.552 & 1.056 & $0.0_{-0.0}^{+0.7}$ \\
  4.918 & $1.0_{-0.8}^{+2.5}$  & 208  & 7.22  & 1.699 & 1.056 & $1.0_{-0.8}^{+2.6}$ \\
  4.951 & $0.0_{-0.0}^{+1.6}$  & 160  & 6.99  & 1.758 & 1.057 & $0.0_{-0.0}^{+2.2}$ \\
  \hline
  \hline
\end{tabular}
\end{table}

\begin{table}[htbp]
\centering
\caption{The results on Born cross sections of $\EE\too\omega\chi_{c2}$ at different energy points. The uncertainties are statistical.}
\label{tab:crosssectionsum2}
\begin{tabular}{ccccccccc}
  \hline
  \hline
  \ \ \ $\sqrt{s}$ (GeV) \ \ \ & \ \ \ $N^{\rm{sig}}$ \ \ \ & \ \ \ $\mathcal{L}_{\rm{int}}$ (pb$^{-1}$) \ \ \ & \ \ \ $\epsilon$ $(\%)$ \ \ \ & \ \ \ 1+$\delta$ \ \ \ & \ \ \ $\frac{1}{|1-\Pi|^{2}}$ \ \ \ & \ \ \ $\sigma^{B}$ (pb) \ \ \ \\
  \hline
  4.358 & $5.3_{-2.2}^{+2.9}$  & 544  & 16.56 & 0.687 & 1.051 & $4.1_{-1.7}^{+2.2}$ \\
  4.377 & $14.1_{-3.6}^{+4.3}$ & 523  & 15.38 & 0.714 & 1.052 & $11.7_{-3.0}^{+3.6}$ \\
  4.387 & $5.0_{-1.9}^{+2.6}$  & 56   & 15.79 & 0.724 & 1.052 & $37.1_{-14.1}^{+19.3}$ \\
  4.396 & $19.0_{-4.2}^{+4.9}$ & 508  & 15.18 & 0.733 & 1.052 & $16.0_{-3.5}^{+4.1}$ \\
  4.416 & $49.8_{-7.3}^{+7.9}$ & 1044 & 15.18 & 0.759 & 1.053 & $19.7_{-2.9}^{+3.1}$ \\
  4.436 & $27.7_{-5.1}^{+5.8}$ & 570  & 14.36 & 0.799 & 1.055 & $20.1_{-3.7}^{+4.2}$ \\
  4.467 & $5.1_{-2.0}^{+2.7}$  & 111  & 13.69 & 0.878 & 1.055 & $18.1_{-7.1}^{+9.6}$ \\
  4.527 & $1.0_{-0.8}^{+2.5}$  & 112  & 11.52 & 1.029 & 1.055 & $3.6_{-2.9}^{+8.9}$ \\
  4.575 & $2.0_{-1.3}^{+2.8}$  & 49   & 10.50 & 1.131 & 1.055 & $16.3_{-10.6}^{+22.8}$ \\
  4.600 & $8.1_{-3.1}^{+3.4}$  & 587  & 10.02 & 1.178 & 1.055 & $5.5_{-2.1}^{+2.3}$ \\
  4.612 & $0.0_{-0.0}^{+1.6}$  & 104  & 9.64  & 1.200 & 1.055 & $0.0_{-0.0}^{+6.3}$ \\
  4.628 & $4.0_{-1.9}^{+3.4}$  & 522  & 9.48  & 1.228 & 1.055 & $3.1_{-1.5}^{+2.7}$ \\
  4.641 & $2.0_{-1.5}^{+3.5}$  & 552  & 9.25  & 1.249 & 1.055 & $1.5_{-1.1}^{+2.6}$ \\
  4.661 & $5.9_{-2.1}^{+2.8}$  & 530  & 9.15  & 1.281 & 1.055 & $4.5_{-1.6}^{+2.1}$ \\
  4.682 & $5.0_{-2.2}^{+3.6}$  & 1669 & 8.96  & 1.313 & 1.055 & $1.2_{-0.5}^{+0.9}$ \\
  4.699 & $0.0_{-1.1}^{+3.3}$  & 536  & 8.65  & 1.338 & 1.055 & $0.0_{-0.8}^{+2.5}$ \\
  4.740 & $0.0_{-0.0}^{+1.6}$  & 164  & 8.48  & 1.394 & 1.055 & $0.0_{-0.0}^{+3.9}$ \\
  4.750 & $0.0_{-0.0}^{+1.6}$  & 367  & 8.43  & 1.408 & 1.055 & $0.0_{-0.0}^{+1.7}$ \\
  4.781 & $1.0_{-0.8}^{+2.5}$  & 513  & 8.28  & 1.447 & 1.055 & $0.8_{-0.6}^{+1.9}$ \\
  4.843 & $1.0_{-0.8}^{+2.5}$  & 527  & 7.97  & 1.519 & 1.056 & $0.7_{-0.6}^{+1.9}$ \\
  4.918 & $1.0_{-0.8}^{+2.5}$  & 208  & 7.63  & 1.598 & 1.056 & $1.9_{-1.5}^{+4.7}$ \\
  4.951 & $0.0_{-0.0}^{+1.6}$  & 160  & 7.40  & 1.630 & 1.057 & $0.0_{-0.0}^{+3.9}$ \\
  \hline
  \hline
\end{tabular}
\end{table}

\begin{table*}[htbp]
\begin{center}
\caption{Systematic uncertainties on cross section measurements at different energy points (in $\%$). The first values in brackets are for $\EE\too\omega\chi_{c1}$, and the second for $\EE\too\omega\chi_{c2}$. }
\label{tab:sumerror}
\begin{tabular}{c c c c c c c c c c c c c c c c c c c c c c c c c c}
  \hline
  \hline
  Source/$\sqrt{s}$ (GeV)  & 4.308 & 4.312 & 4.337 & 4.358 & 4.377 & 4.387 & 4.396 & 4.416 & 4.436 \\
  \hline
  Luminosity & 1.0 & 1.0 & 1.0 & 1.0 & 1.0 & 1.0 & 1.0 & 1.0 & 1.0 \\
  Tracking efficiency & 4.0 & 4.0 & 4.0 & 4.0 & 4.0 & 4.0 & 4.0 & 4.0 & 4.0 \\
  Photon detection & (3.0,-) & (3.0,-) & (3.0,-) & (3.0,3.0) & (3.0,3.0) & (3.0,3.0) & (3.0,3.0) & (3.0,3.0) & (3.0,3.0) \\
  Kinematic fit & (0.2,-) & (0.4,-) & (1.5,-) & (1.3,2.2) & (1.9,1.9) & (2.0,1.8) & (1.4,1.8) & (1.7,1.7) & (2.1,2.2) \\
  $J/\psi$ mass window & (1.6,-) & (1.6,-) & (1.6,-) & (1.6,1.6) & (1.6,1.6) & (1.6,1.6) & (1.6,1.6) & (1.6,1.6) & (1.6,1.6) \\
  Angular distribution & (0.5,-) & (0.4,-) & (0.6,-) & (1.2,0.3) & (2.1,0.7) & (2.2,0.9) & (1.8,0.8) & (2.7,1.5) & (2.6,2.6) \\
  Lineshape & (1.0,-) & (1.9,-) & (1.8,-) & (1.3,1.4) & (2.0,1.3) & (1.3,1.0) & (0.6,1.9) & (1.4,2.7) & (1.3,1.8) \\
  Fit range & (-,-) & (-,-) & (-,-) & (-,1.2) & (-,1.2) & (-,1.2) & (-,1.2) & (-,1.2) & (2.0,1.2) \\
  Signal shape & (-,-) & (-,-) & (-,-) & (-,1.3) & (-,1.3) & (-,1.3) & (-,1.3) & (-,1.3) & (0.2,1.3) \\
  Background shape & (-,-) & (-,-) & (-,-) & (-,2.4) & (-,2.4) & (-,2.4) & (-,2.4) & (-,2.4) & (1.5,2.4) \\
  Branching fraction & (3.0,-) & (3.0,-) & (3.0,-) & (3.0,2.8) & (3.0,2.8) & (3.0,2.8) & (3.0,2.8) & (3.0,2.8) & (3.0,2.8) \\
  Sum & (6.2,-) & (6.4,-) & (6.6,-) & (6.5,7.2) & (7.0,7.2) & (6.9,7.1) & (6.6,7.3) & (7.1,7.6) & (7.5,7.8)\\
  \hline
  \hline
  Source / $\sqrt{s}$ (GeV)  & 4.467 & 4.527 & 4.575 & 4.600 & 4.612 & 4.628 & 4.641 & 4.661 & 4.682 \\
  \hline
  Luminosity & 1.0 & 1.0 & 1.0 & 1.0 & 1.0 & 1.0 & 1.0 & 1.0 & 1.0 \\
  Tracking efficiency & 4.0 & 4.0 & 4.0 & 4.0 & 4.0 & 4.0 & 4.0 & 4.0 & 4.0 \\
  Photon detection & (3.0,3.0) & (3.0,3.0) & (3.0,3.0) & (3.0,3.0) & (3.0,3.0) & (3.0,3.0) & (3.0,3.0) & (3.0,3.0) & (3.0,3.0) \\
  Kinematic fit & (2.3,2.0) & (2.4,2.1) & (1.9,1.7) & (1.8,1.9) & (1.8,2.3) & (2.2,1.8) & (2.2,1.7) & (1.9,2.0) & (1.9,2.2) \\
  $J/\psi$ mass window & (1.6,1.6) & (1.6,1.6) & (1.6,1.6) & (1.6,1.6) & (1.6,1.6) & (1.6,1.6) & (1.6,1.6) & (1.6,1.6) & (1.6,1.6) \\
  Angular distribution & (2.4,2.5) & (3.7,2.9) & (4.1,3.8) & (4.4,4.4) & (4.8,4.9) & (5.0,4.9) & (5.3,4.5) & (5.4,4.2) & (5.6,4.6) \\
  Lineshape & (1.6,1.2) & (1.4,2.1) & (4.8,1.7) & (1.7,0.9) & (2.1,2.0) & (1.3,2.9) & (3.2,2.0) & (1.7,3.0) & (1.0,2.4) \\
  Fit range & (2.0,1.2) & (-,-) & (2.0,-) & (2.0,1.2) & (-,-) & (2.0,-) & (2.0,-) & (2.0,1.2) & (-,-) \\
  Signal shape & (0.2,1.3) & (-,-) & (0.2,-) & (0.2,1.3) & (-,-) & (0.2,-) & (0.2,-) & (0.2,1.3) & (-,-) \\
  Background shape & (1.5,2.4) & (-,-) & (1.5,-) & (1.5,2.4) & (-,-) & (1.5,-) & (1.5,-) & (1.5,2.4) & (-,-) \\
  Branching fraction & (3.0,2.8) & (3.0,2.8) & (3.0,2.8) & (3.0,2.8) & (3.0,2.8) & (3.0,2.8) & (3.0,2.8) & (3.0,2.8) & (3.0,2.8) \\
  Sum & (7.6,7.6) & (7.7,7.3) & (9.3,7.5) & (8.3,8.3) & (8.3,8.4) & (8.7,8.5) & (9.3,8.0) & (8.9,8.7) & (8.6,8.3) \\
  \hline
  \hline
  Source / $\sqrt{s}$ (GeV) & 4.699 & 4.740 & 4.750 & 4.781 & 4.843 & 4.918 & 4.951 \\
  \hline
  Luminosity & 1.0 & 1.0 & 1.0 & 1.0 & 1.0 & 1.0 & 1.0 \\
  Tracking efficiency & 4.0 & 4.0 & 4.0 & 4.0 & 4.0 & 4.0 & 4.0 \\
  Photon detection & (3.0,3.0) & (3.0,3.0) & (3.0,3.0) & (3.0,3.0) & (3.0,3.0) & (3.0,3.0) & (3.0,3.0) \\
  Kinematic fit & (2.1,1.8) & (2.0,2.4) & (2.5,2.0) & (2.4,2.4) & (2.4,3.3) & (2.2,2.3) & (1.9,2.1) \\
  $J/\psi$ mass window & (1.6,1.6) & (1.6,1.6) & (1.6,1.6) & (1.6,1.6) & (1.6,1.6) & (1.6,1.6) & (1.6,1.6) \\
  Angular distribution & (6.0,4.6) & (5.6,5.5) & (6.1,5.6) & (2.4,5.6) & (6.6,5.2) & (6.5,7.5) & (6.9,6.4) \\
  Lineshape & (2.7,2.3) & (2.7,1.2) & (2.6,1.2) & (1.9,1.4) & (1.8,1.3) & (1.9,1.0) & (2.9,2.0) \\
  Fit range & (2.0,-) & (-,-) & (-,-) & (-,-) & (-,-) & (-,-) & (-,-) \\
  Signal shape & (0.2,-) & (-,-) & (-,-) & (-,-) & (-,-) & (-,-) & (-,-)  \\
  Background shape & (1.5,-) & (-,-) & (-,-) & (-,-) & (-,-) & (-,-) & (-,-) \\
  Branching fraction & (3.0,2.8) & (3.0,2.8) & (3.0,2.8) & (3.0,2.8) & (3.0,2.8) & (3.0,2.8) & (3.0,2.8) \\
  Sum & (9.6,8.1) & (9.0,8.6) & (9.4,8.6) & (7.3,8.7) & (9.5,8.7) & (9.4,10.0) & (9.9,9.3) \\
  \hline
  \hline
\end{tabular}
\end{center}
\end{table*}

\clearpage

\end{widetext}

\end{document}